\documentstyle[prd,aps,eqsecnum,preprint,tighten,floats,epsfig]{revtex}

\newcommand{\beq}{\begin{equation}}
\newcommand{\eeq}{\end{equation}}
\newcommand{\beqa}{\begin{eqnarray}}
\newcommand{\eeqa}{\end{eqnarray}}

\newcommand{\kvec}{\bbox k}

\def\elab{E_{\rm lab}}
\def\egcm{E_{\gamma}^{\rm c.m.}}
\def\picm{{\bbox q}_{\pi}^{\rm c.m.}}
\def\thetacm{\theta_{\rm c.m.}}

\begin{document}

\draft

\preprint{\vbox{\hfill SLAC-PUB-8836 \\
          \vbox{\hfill UMN-D-01-5}  \\
          \vbox{\hfill NCSU-TH-01-07}  \\
          \vbox{\hfill NT@UW-00-09}
          \vbox{\vskip0.3in}
          }}

\title{Perturbative QCD and factorization of coherent pion 
photoproduction on the deuteron%
\footnote{\baselineskip=14pt
Work supported by the Department of Energy,
contracts DE-AC03-76SF00515, DE-FG02-98ER41087, DE-FG02-96ER40947,
and DE-FG03-97ER4104.}}

\author{S. J. Brodsky} 
\address{Stanford Linear Accelerator Center,
Stanford University, Stanford, California 94309}

\author{J. R. Hiller} 
\address{Department of Physics,
University of Minnesota-Duluth, Duluth, Minnesota 55812}

\author{Chueng-Ryong Ji} 
\address{Department of Physics,
North Carolina State University, Raleigh, North Carolina 27695}

\author{G. A. Miller}
\address{Department of Physics,
University of Washington, Seattle, Washington 98195}

\date{\today}

\maketitle

\begin{abstract}
We analyze the predictions of perturbative QCD for pion 
photoproduction on the deuteron $\gamma D \to \pi^0 D$ 
at large momentum transfer using the reduced amplitude 
formalism.  The cluster decomposition of the deuteron 
wave function at small binding only allows
the nuclear coherent process to proceed if each nucleon 
absorbs an equal fraction of
the overall momentum transfer.  Furthermore, each nucleon must
scatter while remaining close to its mass shell.  Thus 
the nuclear photoproduction amplitude 
${\cal M}_{\gamma D \to \pi^0 D}(u,t)$ factorizes as a 
product of three factors: (1) the nucleon photoproduction 
amplitude ${\cal M}_{\gamma N_1 \to \pi^0
N_1}(u/4,t/4)$ at half of the overall momentum transfer, 
(2) a nucleon form factor $F_{N_2}(t/4)$
at half the overall momentum transfer, and 
(3) the reduced deuteron form factor $f_d(t)$, which 
according to perturbative QCD, has the same monopole 
falloff as a meson form factor. 
A comparison with the recent JLAB data for 
$\gamma D\rightarrow \pi^0 D$ of 
Meekins {\em et al}.\ [Phys.\ Rev.\ C {\bf 60}, 052201 (1999)] 
and the available $\gamma p\rightarrow \pi^0 p$ 
shows good agreement between the perturbative QCD prediction 
and experiment over a large range of momentum transfers and 
center-of-mass angles. The reduced amplitude prediction
is consistent with the constituent counting rule 
$p^{11}_T {\cal M}_{\gamma D \to \pi^0 D} \to  F(\thetacm)$ 
at large momentum transfer.  This is found to be consistent with 
measurements for photon lab energies $\elab > 3$ GeV at 
$\thetacm=90^\circ$ and $\elab > 10$ GeV at $136^\circ$.
\end{abstract}

\narrowtext

\section{Introduction} \label{sec:intro}

Most phenomena in nuclear physics can be well understood in
terms of effective theories of dynamical nucleons and mesons.  However, in
some cases, conventional approaches to nuclear theory become inadequate,
and the underlying quark and gluon degrees of freedom of nuclei
become manifest.  One such area where QCD makes
testable predictions is exclusive nuclear processes involving high
momentum transfer, such as the elastic lepton-nucleus form factors at
large photon virtuality $q^2$, and scattering reactions such as deuteron
photodisintegration $\gamma D \to p n$ and pion photoproduction 
$\gamma D \to \pi^0 D$ at large transverse momentum.

The predictions of QCD for nuclear reactions are most easily described in
terms of light-cone (LC) wave functions defined at equal LC time
$\tau=t+z/c$~\cite{BPP}.  The deuteron eigenstate can be projected 
on the complete set of baryon number
$B = 2$, isospin $I = 0$, spin $J = 1, J_z = 0, \pm 1$ color-singlet
eigenstates of the free QCD Hamiltonian, beginning with the six-quark Fock
states.  Each Fock state is weighted by an amplitude 
that depends on the LC momentum fractions $x_i=k_i^+/p^+$ and on
the relative transverse momenta $\kvec_{\perp i}$.  There are five
different linear combinations of six color-triplet quarks that make an
overall color singlet, only one of which corresponds to the conventional
proton and neutron three-quark clusters.  Thus, the QCD decomposition
includes four six-quark unconventional states with ``hidden
color"~\cite{JB}.  The spacelike form factors $F_{\lambda
\lambda^\prime}(Q^2)$ measured in elastic lepton-deuteron scattering for
various initial and final deuteron helicities have  exact
representations as overlap integrals of the LC wave functions constructed
in the Drell--Yan--West frame~\cite{DY,W}, where
$q^+ =0$ and $Q^2 = -q^2 = q^2_\perp$.
At large momentum transfer, the leading-twist elastic deuteron form
factors can be written in a factorized form
\beqa \label{factorization}
F_{\lambda \lambda^\prime}(Q^2)
   = \int^1_0 \prod^5_{i=1} d_{x_i}  \int^1_0
          \prod^5_{j=1} d_{y_j}
        \phi_{\lambda^\prime}(x_i,Q)
           T^{\lambda \lambda^\prime}_H(x_i,y_i,Q)
        \phi_\lambda(y_j,Q), 
\eeqa
where the notation $d_{x_i}$ indicates the integral is evaluated
subject to
the condition $\sum_i x_i = 1$,
the $\phi_\lambda(x_i,Q)$ are the deuteron distribution amplitudes,
defined as the integral  of the six-quark LC wave functions
integrated in transverse momentum up to the factorization scale $Q$, and
$T^{\lambda \lambda^\prime}_H$ is the hard scattering amplitude for
scattering six collinear quarks from the initial to final deuteron
directions.  A sum over the contributing color-singlet states is
assumed.  Because the photon and exchanged-gluon couplings conserve the
quark chiralities and the distribution amplitudes project out $L_z = 0$
components of initial and final wave functions, the dominant form factors
at large momentum transfer are hadron-helicity conserving.  The evolution
equation for the distribution amplitudes is given in Refs.~\cite{JB,BJL}.

The hard-scattering amplitude scales as $\left(\alpha_s/Q^2\right)^5$ at
leading order, corresponding to five gluons exchanged among the six
propagating valence quarks.  Higher order diagrams involving
additional gluon exchanges and loops give next-to-leading order
(NLO) corrections of higher order
in $\alpha_s$.  Thus the nominal behavior of the helicity conserving
deuteron form factors is $1/Q^{10}$, modulo the
logarithmic corrections from the running of the QCD coupling and the
anomalous dimensions from the evolution of distribution amplitudes.  In
fact, the measurement~\cite{Alexa} of the high $Q^2 \ge 5$ GeV$^2$ 
helicity-conserving deuteron form factor
$\sqrt{ A(Q^2)}$ appears consistent with the $Q^{10} A(Q^2)$ scaling
predicted by perturbative QCD and constituent
counting rules~\cite{BrodskyFarrar}.

The analogous factorization formulas for deuteron photodisintegration and
pion photoproduction predict the nominal scaling laws 
$s^{11} d\sigma/dt(\gamma D \to n p) \sim $const and 
$s^{13} d\sigma/dt(\gamma D \to \pi^0 D) \sim $const 
at high energies and fixed $\thetacm$.  Comparison with the data
shows this prediction is only successful at the largest
momentum transfers~\cite{Napolitano,Belz,Bochna}.  This is not 
unexpected, since the presence of a large nuclear mass
and a large number of partons involved can 
be expected to delay the onset of leading-twist scaling.

We may use an important simplifying feature of
nuclear dynamics -- the very weak binding of the deuteron state-- to
improve upon the above  discussion.  The
cluster decomposition theorem~\cite{Namyslowski:1980qy}
states that in the zero-binding limit ($B.E. \to 0$), the
LC wave function of the deuteron must reduce to a convolution of
on-shell color-singlet nucleon wave functions
\begin{eqnarray}\label{deuteronwf}
\lim_{B.E. \to 0} \psi^D_{uududd}(x_i,\kvec_{\perp i},\lambda_i) &=&
\int^1_0 dz \int d^2\ell_\perp  \psi^d(z,\ell_\perp)   \nonumber \\
&& \times\psi^p_{uud}(x_i / z,\kvec_{\perp i}+ (x_i/z)\ell_\perp,\lambda_i) \\
&& \times \psi^n_{udd}(x_i / (1-z),\kvec_{\perp i} - [x_i/(1-z)] \ell_\perp,
                     \lambda_i)\,  \nonumber 
\end{eqnarray}
where $ \psi^d(z,\ell_\perp) $ is the reduced ``body'' LC wave function
of the deuteron in terms of its nucleon components.  
Applying this cluster decomposition to an exclusive
process involving the deuteron, one can
derive a corresponding reduced nuclear amplitude 
(RNA)~\cite{BJL,BrodskyJi,BrodskyHiller}. 
Moreover, at zero binding, one may take 
$ \psi^d(z,\ell_\perp) \to
\delta(z -m_p/(m_p + m_n)) \times \delta^{2} (\ell_\perp)$. 
In effect, each nucleon carries half of the deuteron four-momentum.
This approximation is very accurate, because the width of the 
deuteron momentum distribution is less than the
square root of the ratio of the binding energy to the 
nucleon mass, $\approx0.05$.  Furthermore, the deuteron
wave function is vanishingly small for relevant values of the 
momentum transfer, which are of the order of a GeV.  
The deuteron does have Fock-space terms that are $NN\pi$ components,
but the probability of these is very small, and they do not carry much
momentum.  There are $\Delta\Delta$ components, which have an even  
smaller probability, but might carry more momentum. We
shall neglect such effects at the present time, choosing instead to
examine less exotic possibilities. Similar observations can
be made about hidden-color components in the 
deuteron wave function and other six-quark cluster components, 
which we also neglect. In short, our strategy here is to consider 
the most obvious effects first. The general
dominance of the nucleonic part of the Fock space is caused by
the very small binding energy of the deuteron. The other components
enter with relatively large energy denominators.

Thus in the weak nuclear binding limit, the deuteron form factor reduces
to the overlap of nucleon wave functions at half of the momentum
transfer, and $F_D(Q^2) \to f_d(Q^2) F^2_N({Q^2 \over 4})$ where
the reduced form factor $f_d(Q^2)$ is computed from the overlap of the
reduced deuteron
wave functions~\cite{BrodskyJi}.  The reduced deuteron form factor
resembles that of a spin-one meson form factor since its
nucleonic substructure has been factored out.  Perturbative QCD predicts
the nominal scaling
$Q^2 f_d(Q^2) \sim $const~\cite{BJL}.  The measurements
of the deuteron form factor show that this scaling is in fact well
satisfied at spacelike $Q^2 \ge 1$ GeV$^2$~\cite{Alexa}.

\begin{figure}[htbp]
\centerline{\psfig{figure=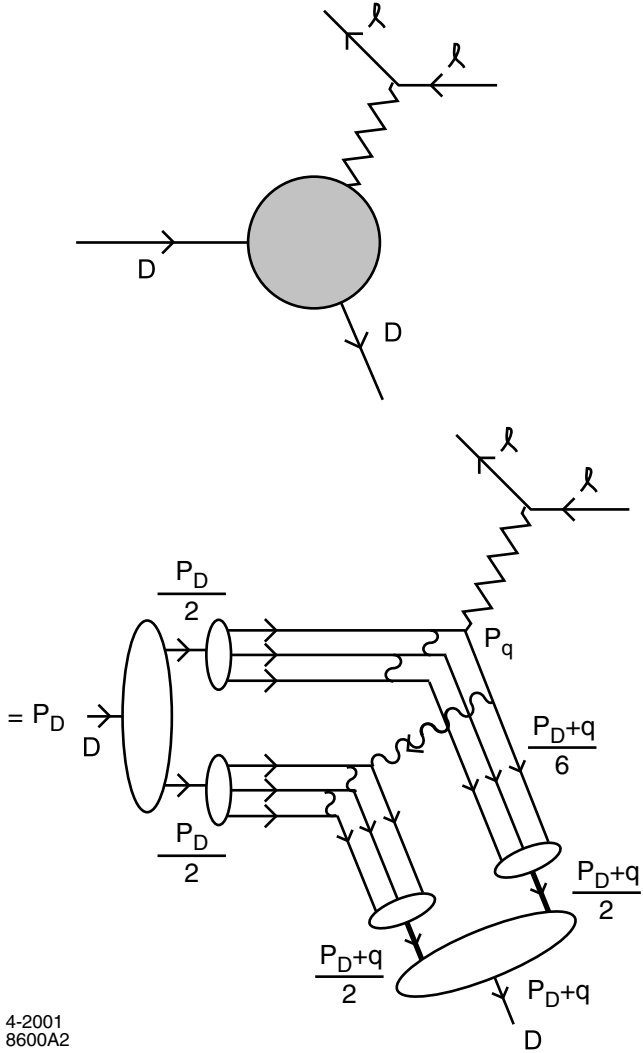}}
\vspace{0.5cm}
\caption{Illustration of the basic QCD mechanism in which
the nuclear amplitude for elastic electron deuteron scattering
$\ell D \to \ell  D$
factorizes as a product of two on-shell nucleon amplitudes.
The propagator of the hard quark line labeled $p_q$ is incorporated into the
reduced form factor $f_d$.
\label{fig:A01}}
\end{figure}

\begin{figure}[htbp]
\centerline{\psfig{figure=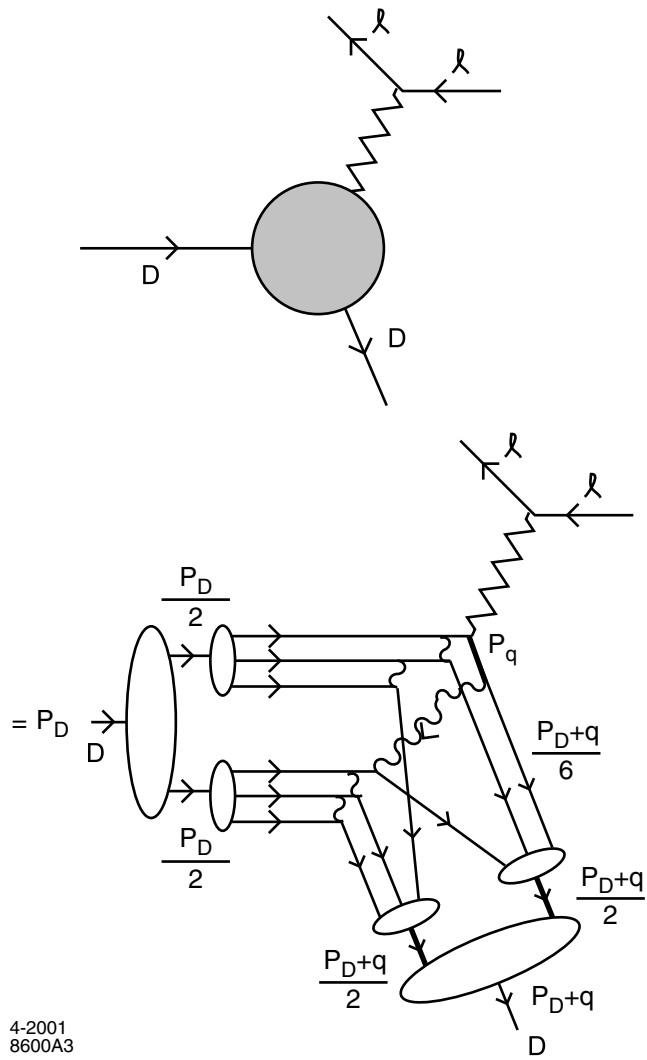}}
\vspace{0.5cm}
\caption{Illustration of the basic QCD mechanism in which
the nuclear amplitude for elastic electron deuteron scattering
$\ell D \to \ell  D$
factorizes as a product of two on-shell nucleon amplitudes.
The quark interchange allows the amplitude to proceed when the
deuteron wave function contains only color-singlet clusters.
\label{fig:A02}}
\end{figure}

The reduced amplitude factorization is evident in the representative QCD
diagram of Fig.~\ref{fig:A01}\@.  Half of the incident photon's
momentum is carried over
to the spectator nucleon by the exchanged gluon.  The struck quark
propagator is off shell with high virtuality $[x_1 (p_D + q) +q/2]^2 \sim
(1+ 2 x_1) q^2/4 \sim q^2/3 $ (using $x_1\sim 1/6$) that provides the
hard scale for the reduced form factor $f_d(Q^2)$.
Figure~\ref{fig:A02}\@ shows a
similar diagram with quark interchange, which is consistent with the
color-singlet clustered structure of the weak-binding amplitude.
Both of these diagrams become independent of the deuteron wave function when
$q$ is greater than about 1 GeV. In this case, one can represent the scattering
amplitude as a product of two factors, one depending on the hard scattering and
the other an integral of the deuteron wave function. The corrections to this
factorization approximation are not computed here, but are estimated
to be of order $1/P_T^2$. 

In this work, we consider a similar analysis of pion photoproduction on the
deuteron $\gamma D \to \pi^0 D$ at weak binding.  The cluster
decomposition of the deuteron wave function at small binding only allows
this process to proceed if each nucleon absorbs an equal fraction of
the overall momentum transfer.  Furthermore, each nucleon must
scatter while remaining close to its mass shell.  Thus we expect the
photoproduction amplitude to factor as
\begin{equation}
{\cal M}_{\gamma D \to \pi^0 D}(u,t) = C(u,t) {\cal M}_{\gamma N_1 \to \pi^0
N_1}(u/4,t/4) F_{N_2}(t/4).
\end{equation}
Note that the on-shell condition requires the center-of-mass angle of pion
photoproduction on the nucleon $N_1$ to be identical to the
center-of-mass angle of pion photoproduction on the deuteron;
the directions of incoming and outgoing particles in the nucleon 
subprocess must be the same as those of the deuteron process.

\begin{figure}[htbp]
\centerline{\psfig{figure=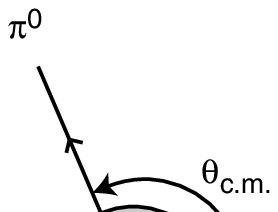}}
\vspace{0.5cm}
\caption{Illustration of the basic QCD mechanism in which
the nuclear amplitude for $\gamma D \to \pi^0 D$
factorizes as a product of two on-shell nucleon amplitudes.
The propagator of the hard quark line labeled $p_q$ is incorporated into the
reduced form factor $f_d$.\label{fig:A03}}
\end{figure}

\begin{figure}[htbp]
\centerline{\psfig{figure=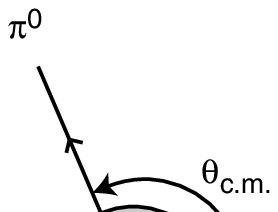}}
\vspace{0.5cm}
\caption{Illustration of the basic QCD mechanism in which
the nuclear amplitude for $\gamma D \to \pi^0 D$
factorizes as a product of two on-shell nucleon amplitudes.
The quark interchange allows the amplitude to proceed when the
deuteron wave function contains only color-singlet clusters.
\label{fig:A04}}
\end{figure}

A representative QCD diagram illustrating the essential features of
pion photoproduction on a deuteron is shown in Fig.~\ref{fig:A03}\@.
The exchanged gluon carries half of the momentum transfer
to the spectator nucleon.  Thus as in the case of the deuteron form factor
the nuclear amplitude contains an extra quark propagator at an
approximate virtuality $t/3$ in addition to the on-shell nucleon
amplitudes.  Thus taking this graph as representative, we can identify
$C(u,t) = C^\prime f_d(t)$, where the constant $C^\prime$ is expected
to be close to unity. 
This correspondence is also shown in Fig.~\ref{fig:A04}\@ that
includes a quark interchange to account for the color-singlet cluster
structure.  This structure predicts the reduced amplitude scaling
\begin{equation}\label{AmplitudeFactored}
{\cal M}_{\gamma D \to \pi^0 D}(u,t) = C^\prime f_d(t) {\cal M}_{\gamma N_1 \to 
\pi^0N_1}(u/4,t/4) F_{N_2}(t/4)\,. \label{goodguy}
\end{equation}
A comparison with elastic electron scattering then yields the following
proportionality of amplitude ratios
\begin{equation}\label{gammaovere}
{{\cal M}_{\gamma D \to \pi^0 D} \over {\cal M}_{e D \to e D}} = C^\prime
{{\cal M}_{\gamma p \to \pi^0 p} \over {\cal M}_{e p \to e p}}.
\end{equation}
More details of the derivation for Eq.~(\ref{AmplitudeFactored}) will be
presented in the following section.  
The normalization is fixed by the requirement that 
this factorization yields the same result as the full counting
rules for ${\cal M}$ in the asymptotic limit. 
Fixing the normalization at a nonasymptotic energy can be a poor
approximation, as can be seen in a recent analysis~\cite{Meekins}.

The new factored form, Eq.~(\ref{AmplitudeFactored}), differs 
significantly from the older 
reduced nuclear amplitude factorization~\cite{BrodskyHiller},
for which
\begin{equation} \label{eq:oldRNA}
{\cal M}_{\gamma D\rightarrow\pi^0 D}^{\rm older}(u,t)\simeq 
   m_{\gamma d\rightarrow\pi^0 d}(u,t)F_N^2(t/4)\,.
\end{equation}
Here $m_{\gamma d\rightarrow\pi^0 d}$ is the reduced amplitude;  it
scales the same as $m_{\gamma \rho \rightarrow\pi^0 \rho}$ at fixed angles 
since  the nucleons of the reduced deuteron $d$ are effectively 
pointlike.  
The advantages of this reduction are that some nonperturbative
physics is included via the nucleon form factors and that 
systematic extension to many nuclear 
processes is possible~\cite{BrodskyHiller}.  The new factorization 
given by Eq.~(\ref{AmplitudeFactored}) is an improvement because 
it includes nonperturbative effects in the pion production process 
itself.  

Recently, JLAB experimental data~\cite{Meekins} on $\pi^0$ 
photoproduction from a deuteron target, up to a photon lab 
energy $\elab=4$ GeV, were presented as an example inconsistent 
with both constituent-counting rules (CCR)~\cite{BrodskyFarrar} 
and RNA~\cite{BrodskyHiller} 
predictions. While the data at $\thetacm=136^\circ$ 
are consistent with the CCR, predicted as $s^{-13}$ scaling for the 
differential cross section $d\sigma/dt$, the data at 
$\thetacm=90^\circ$ exhibit a large disagreement with this 
prediction. Also, the data at both angles were interpreted~\cite{Meekins}
as being inconsistent with the RNA approach.  This is in sharp contrast 
to the recent measurements of the deuteron electric form factor $A(Q^2)$, 
which are consistent with both the CCR and RNA predictions in a similar 
four-momentum transfer range 2 GeV$^2 \le Q^2 \le 6$ 
GeV$^2$~\cite{Alexa}.  

One potential explanation for this disagreement is odderon exchange~\cite{odderon}.  Because the odderon has zero isospin and is
odd under charge conjugation, such an exchange is allowed in the $t$ channel
of $\pi^0$ photoproduction. However, we shall show that
the improved factorization given by Eqs.~(\ref{AmplitudeFactored}) and 
(\ref{gammaovere}) is in reasonably good agreement with  
the recent JLAB data~\cite{Meekins} for $\gamma D\rightarrow \pi^0 D$  
and the available $\gamma p\rightarrow \pi^0 p$ 
data~\cite{Imanishi,Shupe,Bolon,Anderson73,Anderson71}
as well as the existing $e D \rightarrow e D$ and $e p \rightarrow e p$ data.
There is thus no need to invoke any additional anomalous contribution to
understand the new data~\cite{Meekins}. 

We will also predict results for $\gamma D\rightarrow \pi^0 D$ at energies
not yet attained. Furthermore, we will 
analyze the $\pi^0$ transverse momentum 
$P_T$ dependence of the amplitude ${\cal M}_{\gamma D\rightarrow\pi^0 D}$ 
in the c.m.\ frame and find that the scaling of the predicted
${\cal M}_{\gamma D\rightarrow\pi^0 D}$ is not inconsistent with the 
CCR prediction of $P_T^{-11}$ when the photon lab energy is only
a few GeV for both values of $\thetacm$, $90^\circ$ and $136^\circ$.

In the following section, Sec.~\ref{sec:factor}, we first briefly summarize the
kinematics involved in the $\pi^0$ photoproduction process and then present 
a derivation of 
the improved factorization given by Eqs.~(\ref{AmplitudeFactored}) and 
(\ref{gammaovere}).
In Sec.~\ref{sec:results}, we show the numerical results for 
${\cal M}_{\gamma D\rightarrow\pi^0 D}$ as predicted by the factorization 
and compare the results with the recent JLAB data~\cite{Meekins}. We also 
analyze the $P_T$ dependence of ${\cal M}_{\gamma D\rightarrow\pi^0 D}$ 
as a function of $\elab$ for $\thetacm = 90^\circ$ and 
$136^\circ$. 
Our conclusions and some discussion follow in Sec.~\ref{sec:conclusion}.

\section{Kinematics and Factorization} \label{sec:factor}

\subsection{$\pi^0$ Photoproduction kinematics}

The Mandelstam~\cite{Mandelstam} variables of the 
$\gamma D \rightarrow \pi^0 D$ process are given by
\begin{equation} \label{eq:Mandelstam}
s = (q_\gamma + p_D)^2\,,\;\;
t = (q_\gamma-q_\pi)^2\,,\;\;
u = (p_D - q_\pi)^2\,,
\end{equation}
where $p_D = \sum_{a=1}^6 p_a$ is the momentum of the target deuteron
and $q_\gamma$, $q_\pi$, and $p_a$ are the momenta of the photon, pion, 
and $a$th quark of the deuteron.  In the $\gamma$-$D$ c.m.\ frame, 
where experimental results are reported, these variables are related
to the photon energy and pion momentum by
\begin{eqnarray}
s &=& [\egcm + \sqrt{m_D^2 + (\egcm)^2}]^2\,, \nonumber \\
t &=& m_{\pi}^2 - 2 \egcm [\sqrt{m_{\pi}^2 + (\picm)^2} - 
            |\picm|\cos\thetacm]\,, \\
u &=& m_D^2 + m_{\pi}^2 - 2 [ \sqrt{m_D^2 + (\picm)^2}\sqrt{m_{\pi}^2
             +(\picm)^2} + \egcm|\picm|\cos\thetacm]\,, \nonumber 
\end{eqnarray}
with $m_D$ the deuteron mass and $\thetacm$ the angle between 
the photon and the $\pi^0$ in the c.m.\ frame.
Here, also in the c.m.\ frame, the photon energy and
the magnitude of $\pi^0$ momentum are given by 
$\egcm = (s-m_D^2)/2\sqrt{s}$ and 
$|\picm| = \sqrt{(s+m_\pi^2-m_D^2)^2/4s - m_\pi^2}$, respectively.
The transverse momentum of the $\pi^0$ is then given by 
$P_T = |\picm| \sin\thetacm$ and, if all the masses are 
neglected, $P_T \approx \sqrt{tu/s}$.
This simple expression is written to provide qualitative guidance to the
reader. Our numerical calculations use the correctly computed value of
$P_T$.

The Mandelstam variables $s_N$, $t_N$ and $u_N$ of the process
$\gamma N \rightarrow \pi^0 N$ can also be defined, with the deuteron 
momentum in Eq.~(\ref{eq:Mandelstam}) replaced by the nucleon momentum 
$p_N = \sum_{a=1}^3 p_a$. In the $\gamma$-$N$ c.m.\ frame of the
$\gamma N \rightarrow \pi^0 N$ process, the photon energy  
and the magnitude of the $\pi^0$ momentum are given by 
$(\egcm)_N = (s_N-m_N^2)/2\sqrt{s_N}$ and 
$(\picm)_N = \sqrt{(s_N+m_\pi^2-m_N^2)^2/4 s_N - m_\pi^2}$,
respectively, with the nucleon mass being $m_N$. 

One can find the magnitude of the invariant amplitude 
${\cal M}_{\gamma D\rightarrow\pi^0 D}(u,t)$ 
from the experimental differential cross section data  by using
\begin{equation} \label{eq:direct}
|{\cal M}_{\gamma D\rightarrow\pi^0 D}(u,t)| = 
   4 (s-m_D^2) \sqrt{\pi \frac{d\sigma}{dt}(\gamma D \rightarrow \pi^0 
D)}\,. 
\end{equation}
Similarly, we obtain the invariant amplitude 
$|{\cal M}_{\gamma N\rightarrow\pi^0 N}(u_N,t_N)|$
from the available data for 
$\gamma p \rightarrow \pi^0 
p$~\cite{Imanishi,Shupe,Bolon,Anderson73,Anderson71}.
The proton data and the factorization formula (\ref{AmplitudeFactored}) can
then be used to predict $|{\cal M}_{\gamma D\rightarrow\pi^0 D}(u,t)|$.
We take the generic nucleon form factor $F_N (t)$ to be
$\left[ 1 - t/(0.71\,\mbox{GeV}^2) \right]^{-2}$ and 
the reduced deuteron form factor~\cite{BrodskyChertok}  
$f_d (t) \approx 2.14/ \left[ 1 - t/(0.28\,\mbox{GeV}^2) \right]$, 
as determined by the analyses of 
the elastic deuteron form factors~\cite{Arnold}.
As one can see in the previous analysis~\cite{BJL}, the experimental
data for $|t| \le 2\,\mbox{GeV}^2$ are better described without the 
logarithmic corrections. The normalization constant $C^\prime$ 
in Eq.~(\ref{AmplitudeFactored}) is then fixed by 
the largest $\elab$ data point of $\gamma D \rightarrow \pi^0 D$ 
amplitude~\cite{Meekins} at $\thetacm = 90^\circ$ and is obtained as 
$C^\prime \approx 0.8$.

\subsection{New improved RNA factorization}

The factorization given by Eq.~(\ref{AmplitudeFactored}) can be derived
in analogy with earlier work on the deuteron form 
factor~\cite{BrodskyJi}.  The first step is to replace 
at the quark level the electromagnetic vertex 
$\gamma q\rightarrow q$ with a photoproduction amplitude 
$\gamma q\rightarrow \pi^0 q$.
Let ${\cal M}_{\gamma q\rightarrow\pi^0 q}(q_\gamma,q_\pi,p_a)$ 
be the amplitude for this subprocess, with $q_\gamma$, $q_\pi$, and $p_a$ 
the momenta of the photon, pion, and $a$th quark, respectively.  The full 
amplitude for the hadronic process $\gamma D\rightarrow \pi^0 D$ can be 
transcribed from Eq.~(2.10) of~\cite{BrodskyJi}, with insertion of 
${\cal M}_{\gamma q\rightarrow\pi^0 q}$, as
\begin{equation}  \label{eq:Amplitude}
{\cal M}_{\gamma D\rightarrow\pi^0 D}(u,t)=\sum_{a=1}^6\int 
[dx]_i[d^2k_\perp]_i
  \Psi_D^*(x_i,{\bbox k}_{\perp i}+(\delta_{ia}-x_i){\bbox q}_\perp)
          {\cal M}_{\gamma q\rightarrow\pi^0 q}(q_\gamma,q_\pi,p_a)
                                    \Psi_D(x_i,{\bbox k}_{\perp i})
\end{equation}
where $\Psi_D$ is the valence wave function, $q\equiv q_\gamma-q_\pi$
is the momentum transfer, and
\begin{equation}
[dx]_i = \delta(1-\sum_{i=1}^6 x_i)\prod_{i=1}^6\frac{dx_i}{x_i}\,, \;\;
{[d^2k_\perp]_i} = 16\pi^3\delta^2(\sum_{i=1}^6{\bbox k}_{\perp i})
    \prod_{i=1}^6\frac{d^2k_{\perp i}}{16\pi^3}\,. 
\end{equation}
The reference frame has been chosen such that $q^+\equiv q^0+q^3=0$.

The deuteron wave function factorizes in the manner described by
Eq.~(2.23) of~\cite{BrodskyJi}, which reads
\begin{eqnarray}
\sum_{a=1}^6
  \Psi_D^*(x_i,{\bbox k}_{\perp i}+(\delta_{ia}-x_i){\bbox q}_{\perp})
  &=&\left[ \sum_{a=1}^3\sum_{b=4}^6+\sum_{a=4}^6\sum_{b=1}^3\right]
   \frac{x_a}{1-x_a}\frac{1}{q_\perp^2} \\
&& \times
   V(x_i,(\delta_{ia}-x_i){\bbox q}_{\perp};
     x_j,[y\delta_{ja}+(1-y)\delta_{jb}-x_j]{\bbox q}_{\perp}) 
   \nonumber \\
&&\times
 \psi_N(z_i,{\bbox k}_{\perp i}^{\,\prime}+
              (\delta_{ia}-z_i)y{\bbox q}_{\perp})
\nonumber \\
&& \times
 \psi_N(z_j,{\bbox k}_{\perp j}^{\,\prime}+
             (\delta_{jb}-z_j)(1-y){\bbox q}_{\perp}) \psi^d(0)\,,  
  \nonumber
\end{eqnarray}
where $\psi^d(0)$ is the body wave function of the deuteron at the
origin and
\begin{equation}
y=\sum_{i=1}^3 x_i\,,\;\;
{\bbox l}_\perp=\sum_{i=1}^3 {\bbox k}_{\perp i}\,,\;\;
z_i=\frac{x_i}{y}\,,\;\;
{\bbox k}_{\perp i}^{\,\prime}={\bbox k}_{\perp i}-z_i{\bbox l}_\perp\,,\;\;
z_j=\frac{x_j}{1-y}\,,\;\;
{\bbox k}_{\perp j}^{\,\prime}={\bbox k}_{\perp j}+z_j{\bbox l}_\perp\,.
\end{equation}
In the weak binding limit, the value of $y$ is approximately 1/2,
${\bbox l}_\perp$ is approximately zero, and the kernel $V$ contributes
only a constant.  The deuteron amplitude (\ref{eq:Amplitude}) reduces to 
the analog of Eq.~(2.24) in Ref.~\cite{BrodskyJi}
\begin{eqnarray}
{\cal M}_{\gamma D\rightarrow\pi^0 D}(u,t)&=&
                       \frac{C}{q_\perp^2}|\psi^d(0)|^2
\left\{\sum_{a=1}^3\int [dz]_i[d^2k'_\perp]_i   \right.   \nonumber  \\
&& \times \psi_N^*\left(z_i,{\bbox k}_{\perp i}^{\,\prime}
                           +(\delta_{ia}-z_i)\frac{{\bbox q}_\perp}{2}\right)
      {\cal M}_{\gamma q\rightarrow\pi^0 q}(q_\gamma,q_\pi,p_a)
      \psi_N(z_i,{\bbox k}_{\perp i}^{\,\prime})          \\
&& \left. \times
\sum_{b=4}^6\int [dz]_j[d^2k'_\perp]_j
      \psi_N^*\left(z_j,{\bbox k}_{\perp j}^{\,\prime}
                           +(\delta_{jb}-z_j)\frac{{\bbox q}_\perp}{2}\right)
      \psi_N(z_j,{\bbox k}_{\perp j}^{\,\prime}) + (a\leftrightarrow 
b)\right\}\,.
\nonumber  
\end{eqnarray}

This result does not quite factorize because the quark amplitude 
${\cal M}_{\gamma q\rightarrow\pi^0 q}$ depends on the full $q_\gamma$ 
and $q_\pi$, 
whereas the individual nucleons experience momentum transfers of 
$(q_\gamma-q_\pi)/2$.  
To relate this quark amplitude to the one for a subprocess involving only 
a nucleon, we use the spin-averaged form of the amplitude obtained by 
Carlson and Wakely~\cite{CarlsonWakely}
\begin{equation}
|{\cal M}_{\gamma q\rightarrow\pi^0 q}(\hat{u},\hat{t})|^2 \sim 
   \hat{t}\frac{\hat{s}^2+\hat{u}^2}{\hat{s}^2\hat{u}^2}\,,
\label{cw}\end{equation}
where $\hat{s}=(q_\gamma+p_a)^2$, $\hat{t}=(q_\gamma-q_\pi)^2=t$, and 
$\hat{u}=(p_a-q_\pi)^2$.  For photoproduction from a single nucleon,
embedded in the deuteron, we have instead the quark-level invariants
\begin{equation}
\hat{s}_N=(q_\gamma/2+p_a)^2\,,\;\; 
\hat{t}_N=(q_\gamma/2-q_\pi/2)^2=\hat{t}/4\,,\;\;
\hat{u}_N=(p_a-q_\pi/2)^2\,.
\end{equation}
The quark momentum is the same in both cases simply because it is
the same quark.  In the zero-mass limit we have $\hat{s}_N=\hat{s}/2$ and
$\hat{u}_N=\hat{u}/2$.  This [with Eq.~(\ref{cw})] leaves 
$|{\cal M}_{\gamma q\rightarrow\pi^0 q}(\hat{u},\hat{t})|^2\simeq 
    |{\cal M}_{\gamma q\rightarrow\pi^0 q}(\hat{u}_N,\hat{t}_N)|^2$.
Furthermore, the above values of $\hat s_N, \hat t_N,\hat u_N$
correspond to using $q_\gamma/2$ and $q_\pi/2$ in evaluating the
proton photoproduction amplitude. With these values 
the factorization can now be completed to obtain 
Eq.~(\ref{AmplitudeFactored}). A similar derivation can
be constructed 
for a relation between the amplitudes of $e D \rightarrow e D $ and 
$e N \rightarrow e N$ processes, from
which one can prove Eq.~(\ref{gammaovere}).
The corrections to these factorization formulas are generally
expected to be of order $1/P_T^2$. The computation of these terms is beyond
the scope of this paper.

\section{Comparison With Experiment} \label{sec:results}

\begin{figure}[htbp]
\centerline{\psfig{figure=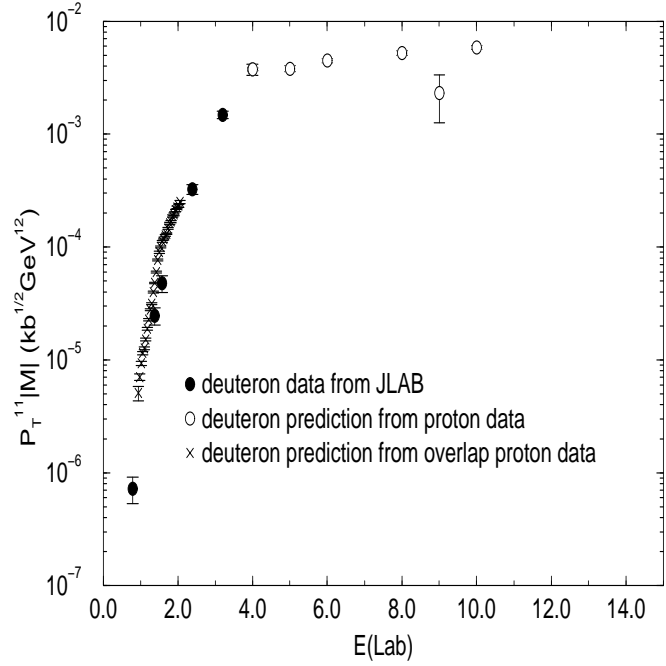,height=10cm,width=10cm}}
\vspace{0.5cm}
\caption{$P_T^{11}|{\cal M}_{\gamma D\rightarrow\pi^0 D}|$ vs 
photon lab energy $\elab$ at $\thetacm = 90^\circ$.  The filled 
circles are obtained directly from the recent JLAB 
$\gamma D \rightarrow \pi^0 D$ data~\protect\cite{Meekins}, and the 
crosses and open circles are our predictions from the 
$\gamma p \rightarrow \pi^0 p$ data presented 
in Refs.~\protect\cite{Imanishi,Shupe,Bolon}.
Note that an open circle is overlaid on top of a filled circle
for the overlapping data point at $\elab = 4$ GeV.
\label{fig:pt11amp90}}
\end{figure}

\begin{figure}[htbp]
\centerline{\psfig{figure=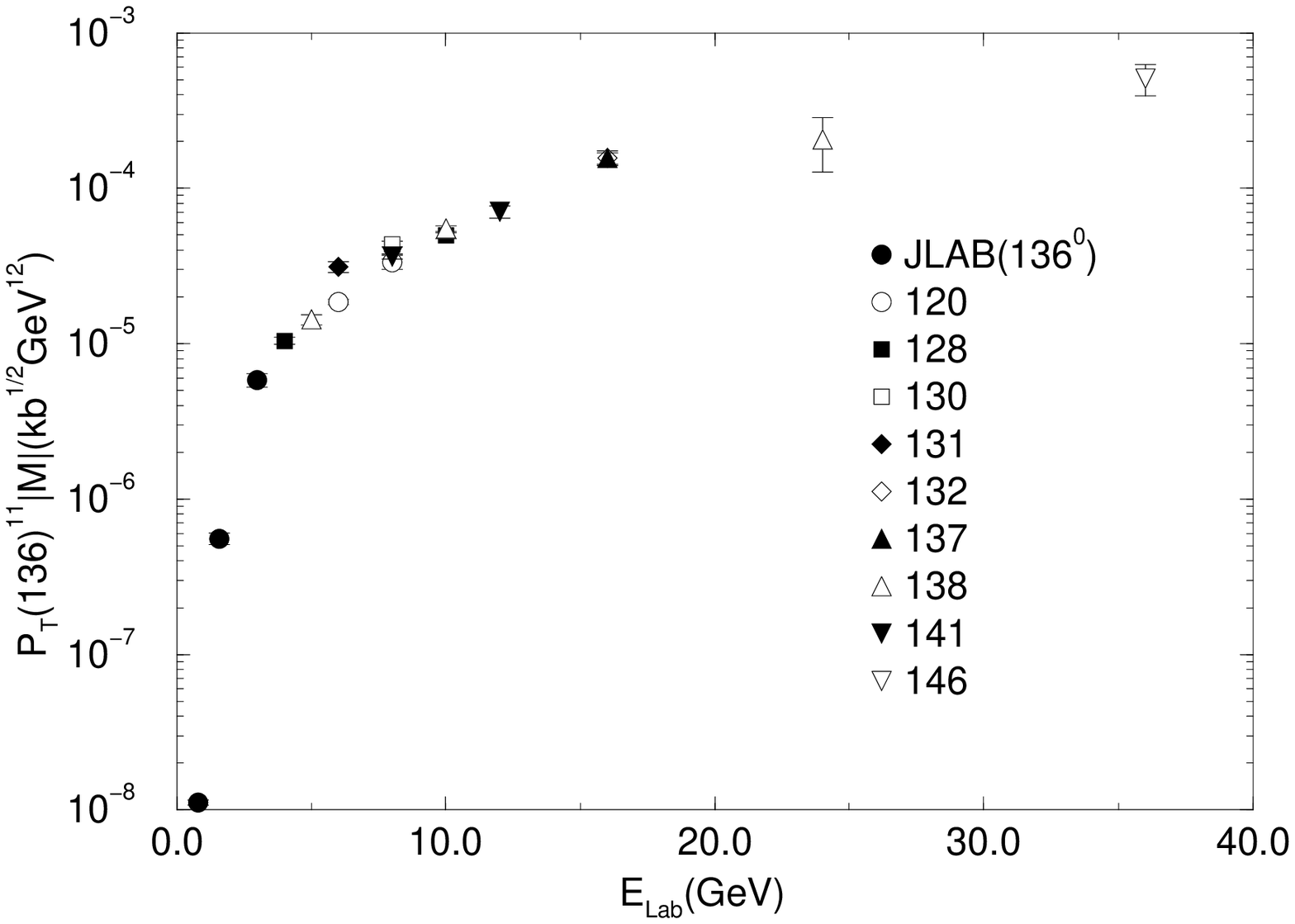,height=10cm,width=10cm}}
\vspace{0.5cm}
\caption{$P_T^{11}(136)|{\cal M}_{\gamma D\rightarrow\pi^0 D}|$ vs 
photon lab energy $\elab$ with $P_T^{11}(136)$ defined by the $P_T$ value 
computed at $\thetacm=136^\circ$. This definition is due to the fact 
that the $\thetacm$ values of the $\gamma p \rightarrow \pi^0 p$ data 
are near $136^\circ$ but not exactly equal.  The filled circles are 
obtained directly from the recent JLAB $\gamma D \rightarrow \pi^0 D$ 
data~\protect\cite{Meekins}, and the other symbols are our predictions 
based on $\gamma p \rightarrow \pi^0 p$ data at angles near
$\thetacm=136^\circ$, presented in 
Refs.~\protect\cite{Imanishi,Shupe,Bolon}. 
\label{fig:pt11amp136}} 
\end{figure}

From the recent JLAB $\gamma D \rightarrow \pi^0 D$ data~\cite{Meekins}, 
we computed the corresponding invariant amplitudes using 
Eq.~(\ref{eq:direct}) both for $\thetacm = 90^\circ$ and $136^\circ$. 
We then used our factorization formula Eq.~(\ref{AmplitudeFactored}) 
to predict $|{\cal M}_{\gamma D\rightarrow\pi^0 D}|$ with input 
from the available $\gamma p \rightarrow \pi^0 p$ 
data~\cite{Imanishi,Shupe,Bolon,Anderson73,Anderson71}.  The results are 
presented in Figs.~\ref{fig:pt11amp90} and \ref{fig:pt11amp136}.
There are no data for pion production from a neutron target, and we assume
here that the proton and neutron amplitudes have the same dependence on
$u$ and $t$.  This assumption is reasonable for the production of neutral 
pions, which contain $u$ and $d$ quarks with equal probability and
therefore couple to the proton and neutron in similar ways.  We do not
consider charged pion production.

In Fig.~\ref{fig:pt11amp90}, the normalization of our prediction is 
fixed (at $C'=0.8$)
by the overlapping data point at $\elab = 4$ GeV, which is the 
highest photon lab energy used in the JLAB $\gamma D \rightarrow \pi^0 D$ 
experiment~\cite{Meekins}.  It is interesting to find that the general 
trend of our prediction (the open circles) is very similar to that of the 
direct result from the JLAB data~\cite{Meekins}, shown as filled circles. 
The prediction is remarkably consistent with the CCR prediction.
In addition, our ``prediction'' in the $\elab$ overlap region, denoted by 
crosses, mimics the shape of the direct result.  The crosses are 
systematically above all the filled circles by 50\% or more (on a 
linear scale).  This difference could be absorbed into the determination 
of the normalization; however, the factorization is expected to be less 
accurate at these lower energies.
We note that the virtuality of the struck-quark propagator discussed
in Sec.~\ref{sec:intro} (see also Figs.~\ref{fig:A03} and 
\ref{fig:A04}) is approximately given by $|t|/3\approx 2$ GeV$^2$ 
for $\elab = 4$ GeV, which satisfies the condition mentioned in the 
Introduction.  We therefore fix the normalization $C^\prime$ of our 
factorization formula Eq.~(\ref{goodguy}) from the data at the
highest available photon lab energy, {\em i.e.} $\elab = 4$ GeV.
We expect non-negligible corrections to our formula at all energies
we consider here, but these are expected to become increasingly small,
falling off as $P_T^{-2}$, as the energy increases.
Also, one should note that there is a resonance contribution in 
the $\gamma p \rightarrow \pi^0 p$ data~\cite{Imanishi}, in the region 
of 700 MeV $\le \elab \le$ 800 MeV, which could bias a normalization
done at lower energies. 

In obtaining Fig.~\ref{fig:pt11amp136}, we use input from the proton 
data~\cite{Imanishi,Shupe,Bolon,Anderson73,Anderson71} in the vicinity
of $\thetacm=136^\circ$ to compute the scattering amplitude
for a deuteron target. We were unable to find
$\gamma p \rightarrow \pi^0 p$ data exactly at 
$\thetacm = 136^\circ$ in the $\elab$ energy range considered
here.  Using the same procedure for determining
the normalization as in obtaining Fig.~\ref{fig:pt11amp90}, 
we find that our prediction is nicely connected to the direct calculation 
from the JLAB data~\cite{Meekins}, again shown as filled circles.
The prefactor $P_T^{11}$ is computed at 
$\thetacm=136^\circ$ for all data points.  
The scaling behavior predicted by 
perturbative QCD, $P_T^{11}|{\cal M}_{\gamma D\rightarrow\pi^0 D}|
\sim$ const at fixed $\thetacm$, does not appear to work as well at 
$\thetacm=136^\circ$ in Fig.~\ref{fig:pt11amp136} in comparison with 
the $\thetacm=90^\circ$ data shown in Fig.~\ref{fig:pt11amp90}\@.  
This is in part due to the fact that the high 
power of 11 makes the prefactor $P_T^{11}$ very sensitive
to the variation in the values of $\thetacm$. The accuracy of the
present calculations is therefore limited by the lack of
data for both the deuteron and proton target at the 
same fixed values of $\theta$, and more $\gamma D \rightarrow \pi^0 D$ 
and $\gamma p \rightarrow \pi^0 p$ data are needed.
However, if we look only at the data points that are close
to $\thetacm = 136^\circ$, such as the filled triangles at
$137^\circ$ and open triangles at $138^\circ$, and take
experimental errors into account, 
CCR scaling is not inconsistent for $\elab$ above 10 GeV.

\section{Conclusions and Discussion} \label{sec:conclusion}

We have analyzed the predictions of perturbative QCD for coherent
photoproduction on the deuteron $\gamma D \to \pi^0 D$ 
at large momentum transfer using a new form of reduced
amplitude factorization displayed in Eq.~(\ref{goodguy}).
The underlying principle of the analysis is the cluster
decomposition theorem for the deuteron wave function at 
small binding: the nuclear coherent process can proceed 
only if each nucleon absorbs an equal fraction of
the overall momentum transfer.  Furthermore, each nucleon must
scatter while remaining close to its mass shell.  Thus  the nuclear 
photoproduction amplitude ${\cal M}_{\gamma D \to \pi^0 D}(u,t)$ 
factorizes as a product of three factors: (1) the nucleon photoproduction 
amplitude ${\cal M}_{\gamma N_1 \to \pi^0
N_1}(u/4,t/4)$ at half of the overall momentum transfer and at the 
same overall center-of-mass angle, (2) a nucleon form factor $F_{N_2}(t/4)$
at half the overall momentum transfer, and (3) the reduced deuteron 
form factor $f_d(t)$, which according to perturbative QCD, has the 
same monopole falloff as a meson form factor. 
The on-shell condition requires the center-of-mass angle of pion
photoproduction on the nucleon $N_1$ to be commensurate with the
center-of-mass angle of pion photoproduction on the deuteron.  The 
reduced amplitude prediction
is consistent with the constituent counting rule 
$p^{11}_T {\cal M}_{\gamma D \to \pi^0 D} \to  F(\thetacm)$ 
at large momentum transfer. 
A comparison with the recent JLAB data for $\gamma D\rightarrow \pi^0 D$  
of Meekins {\em et al}.~\cite{Meekins} and the available 
$\gamma p\rightarrow \pi^0 p$
data~\cite{Imanishi,Shupe,Bolon,Anderson73,Anderson71}
shows good qualitative agreement between the perturbative 
QCD prediction and experiment over a large range of momentum 
transfers and center-of-mass angles.  It would be useful
to confirm  or deny this  agreement with further measurements
on both proton and deuteron targets. However, there are additional
uncertainties due to the lack of knowledge about pion production from the
neutron.

We have also used reduced amplitude scaling 
for the elastic electron-deuteron scattering to show
\begin{equation}\label{gammaovere2}
{{\cal M}_{\gamma D \to \pi^0 D} \over {\cal M}_{e D \to e D}} = C^\prime
{{\cal M}_{\gamma p \to \pi^0 p} \over {\cal M}_{e p \to e p}}.
\end{equation}
This scaling is also consistent with experiment.  
The constant $C^\prime$ is found to be close to 1, suggesting 
similar underlying hard-scattering contributions. No anomalous 
contributions such as might derive from odderon exchange are required.

\section*{Acknowledgments}
This work was supported in part by the Department of Energy,
contracts DE-AC03-76SF00515 (S.J.B.),  DE-FG02-98ER41087 (J.R.H.), 
DE-FG02-96ER40947 (C.-R.J.), and DE-FG03-97ER4104 (G.A.M.). 


\end{document}